\documentclass[aps,prd,preprint,superscriptaddress,tightenlines,nofootinbib]{revtex4}

\usepackage{graphicx}
\usepackage{dcolumn}
\usepackage{bm}

\def\ebeam{E_{\mathrm{beam}}}
\def\roots{\sqrt s}
\def\myratio{{\cal R}_{\tau\tau}^{\Upsilon} }
\def\myratioexp{{\cal R}_{\tau\tau}^{\Upsilon \, {\mathrm{th}}} }
\def\myratiooff{{\cal R}_{\tau\tau}^{\mathrm{Off}}}
\def\myratiooffexp{{\cal R}_{\tau\tau}^{\mathrm{Off} \, {\mathrm{th}}}}
\def\etal{{\it et~al.}}

\begin{document}

\preprint{CLNS 06/1961}       
\preprint{CLEO 06-10}         

\title{
First Observation of $ \Upsilon(3S)\to \tau^+\tau^-$ and Tests of Lepton
 Universality in $ \Upsilon$ Decays.
}



\author{D.~Besson}
\affiliation{University of Kansas, Lawrence, Kansas 66045}
\author{T.~K.~Pedlar}
\affiliation{Luther College, Decorah, Iowa 52101}
\author{D.~Cronin-Hennessy}
\author{K.~Y.~Gao}
\author{D.~T.~Gong}
\author{J.~Hietala}
\author{Y.~Kubota}
\author{T.~Klein}
\author{B.~W.~Lang}
\author{R.~Poling}
\author{A.~W.~Scott}
\author{A.~Smith}
\author{P.~Zweber}
\affiliation{University of Minnesota, Minneapolis, Minnesota 55455}
% ...
\collaboration{CLEO Collaboration} 
\noaffiliation


\date{July 12, 2006}

\begin{abstract} 
Using data collected with the CLEOIII detector at the CESR 
 $e^+e^-$ collider,
 we report on a first observation of the decay $\Upsilon(3S) \to \tau^+\tau^-$,
 and precisely measure the ratio of branching fractions of 
 $\Upsilon(nS),n=1,2,3,$  to $\tau^+\tau^-$ and $\mu^+\mu^-$ final states,
 finding agreement with expectations from lepton universality.
 We derive
 absolute branching fractions for these decays, and also set a limit on
 the influence of a low mass CP-odd Higgs boson in the decay of 
 the $\Upsilon(1S)$.
\end{abstract}

\maketitle


 In the Standard Model,  the
 couplings between leptons and gauge bosons are independent of the lepton flavor, so
  the branching fractions for the decay $\Upsilon(nS)\to {\it l^+}{\it 
l^-} $ should be
   independent of the flavor of the lepton ${\it l}$, 
except for negligible lepton mass effects.
Any deviation from unity for
    the ratio of branching fractions
     $ \myratio = B(\Upsilon(nS) \to \tau^+\tau^-)/B(\Upsilon(nS) 
\to \mu^+\mu^-) $ 
 would indicate the presence of new physics.
The ratio $\myratio$ is sensitive to the mechanism proposed in \cite{miguel}, 
in which a low mass CP-odd Higgs boson, $A^0$,
mediates the  decay chain $\Upsilon(1S) \to \eta_b \gamma, \eta_b \to A^0 \to
 \tau^+\tau^-$.

CLEO has recently measured the partial width, $\Gamma_{ee}$, from
 $e^+e^- \to \Upsilon(nS)$~\cite{pivarski}, 
 $n=1,2,3$, as well as the branching fraction for $\Upsilon(nS) 
\to \mu^+\mu^-$~\cite{idanko}.
  This analysis complements these measurements by measuring 
  $\myratio$ directly, and scales this result to obtain 
$B(\Upsilon(nS) \to \tau^+\tau^-)$.
An upper limit on the product branching fraction 
 $B( \Upsilon(1S) \to \eta_b \gamma) B( \eta_b \to A^0 \to \tau^+\tau^- )$ 
is extracted.

The data in this analysis were obtained with the CLEOIII  
detector~\cite{CLEOdet,CLEOTwodet} at the symmetric $e^+e^-$ collider CESR. 
The detector includes a precision 
tracking system in a solenoidal magnetic field, a CsI calorimeter, a
 Ring Imaging Cherenkov detector, and muon chambers.
The data are identical to those in \cite{idanko}, 
including both on-resonance and off-resonance subsamples at the 
$\Upsilon(nS), n=1,2,3$.
In addition, $ 0.5 \,\mathrm{fb}^{-1}$ at the $\Upsilon(4S)$
and $0.6 \,\mathrm{fb}^{-1}$ 40 MeV below the $\Upsilon(4S)$ peak were used as 
 control samples. 

The analysis technique, similar to that  in \cite{idanko}, isolates 
the $\Upsilon \to \mu^+\mu^- , \tau^+\tau^-$ signals by subtracting a scaled
number of events observed off-resonance 
from the number observed in 
on-resonance data, and, after further background correction, attributes the 
remaining signal to $\Upsilon \to {\it l}^+{\it l}^-$.
Selection criteria are developed 
to isolate $\mu^+\mu^-$ and $\tau^+\tau^-$ final states using a subset 
of the data acquired near the $\Upsilon(4S)$. 
Being dominated by strong interaction decays, the full width of the 
$\Upsilon(4S)$ is sufficiently large that $\Upsilon(4S) \to \tau^+\tau^-$ 
decays in our sample are negligible.
Another subset of 
 $\Upsilon(4S)$ data is used to verify
that the scaled subtraction
produces no signal for $\Upsilon(4S)\to \tau^+\tau^-, \mu^+\mu^-$,
indicating  that non-$\Upsilon$ backgrounds are suppressed by the subtraction. 
Furthermore, the  off-resonance
 production cross-sections for $\tau^+\tau^-$ and $\mu^+\mu^-$ are verified to
 agree with theoretical expectations.

The 
final states chosen for both the $\Upsilon \to \mu^+\mu^-$ and $\Upsilon \to 
\tau^+\tau^-$ decays are required to have exactly two good quality charged 
tracks of opposite charge. In this Letter, $\roots$ denotes the
 total available center of mass energy, $X_P$ is a particle's
 momentum, $P$,  scaled to the beam energy, $\ebeam$. The energy deposited 
by a  particle in the calorimeter is  $E_{\mathrm{CC}}$.

Selection criteria for the $\mu^+\mu^-$ final state closely follow those 
of \cite{idanko}, requiring tracks with $0.7 <X_P < 1.15 $, of which at 
least one is 
positively identified  as a muon, using
both the muon chambers and the calorimeter. The opening angle between
 the associated momenta must be greater than $170^\circ$.
Events must have no more than one isolated shower that is unassociated with  a
track and has energy greater than $1\%$ of the beam energy. 

At least two neutrinos from final states of $e^+e^- \to \tau^+\tau^-$ escape detection.
Thanks to CLEOIII's hermeticity, the following criteria select such events despite
 the energy carried away by the unreconstructed neutrinos.
 The total charged-track momentum transverse to the beam direction must be greater
than 10\% of $\ebeam$, and the total charged-track momentum must point into the barrel region
 of the detector where tracking and calorimetry are optimal. Events with 
collinear tracks are eliminated. Tracks are required to have $0.1<X_P<0.9$, 
 where the upper limit is chosen to minimize pollution from two-particle 
final states.
The total observed energy  due to 
charged and neutral particles in the calorimeter is similarly required to be 
between $0.2 \roots$ and $0.9 \roots$. To reduce 
overlap confusion between neutral and charged particles, tracks and associated
 showers must satisfy $E_{\mathrm{CC}}/P<1.1 $.
Potential backgrounds due to cosmic rays are accounted for 
as in \cite{idanko}. In all cases these were negligible.

Final states are further exclusively 
divided according to the results of particle identification into $(e,e)$, 
$(\mu,e)$, $(e,\mu)$, $(\mu,\mu)$, $(e,X)$, $(\mu,X)$, $(X,X)$ sub-samples, 
with
  particles listed in descending momentum order.
 The first (second) particle listed
 is referred to as the tag (signal).
 Lepton identification requires $P > 500 \, \mathrm{MeV}$
 to ensure that the track intersects the calorimeter.
Electrons are identified by requiring that  $0.85 < E_{\mathrm{CC}}/P < 1.10$, 
and that the specific ionization along the track's path in the drift 
chamber be consistent with the  expectation for an electron.
A muon candidate in $\tau$ decays is a charged track which is not identified as an 
electron, having $P > 2 \,\mathrm{GeV}/c$ ($ > 1.5 \,\mathrm{GeV}/c$) 
for a tag (signal) track
  and confined to the central barrel 
where beam related background is a minimum.
 Furthermore, the muon candidate must 
penetrate at least three interaction lengths into the muon detector,
 and satisfy $100 \,\mathrm{MeV} <  E_{\mathrm{CC}} < 600 \,\mathrm{MeV}$.
 The missing momentum and energy criteria ensure
 that virtually all pions mistakenly identified as muons come
 from $\tau$ pair decays and 
 are properly accounted for in the simulation, as shown by the agreement
 of off-resonance cross sections with expectations.
Particles identified as neither e nor $\mu$ are designated $X$, and are 
a mixture of hadrons and unidentified leptons.

The decay products of $\tau$ pairs from $\Upsilon$ decays 
tend to be separated into distinct hemispheres.
Since the photon spectrum expected in $\tau$ decays depends 
on the identity of the charged particle, calorimeter showers are 
assigned to either the tag or signal hemisphere according to 
their proximity to the tag side track direction. This separation
 into distinct hemispheres is not perfect, so there is some correlation
 between the photon spectrum on tag and signal sides of the event.
 The  modes with two identified leptons each require fewer than 6 showers and
a maximum shower energy  below $ 0.1 \, \ebeam$.
Defining $ E_{\rm neut}^{\rm tag}$ ( $ E_{\rm neut}^{\rm sig}$ ) as
 the total isolated shower energy in the calorimeter 
 on  the tag (signal) sides of the event, 
the $(e,e)$ mode assignment requires
  $ E_{\rm neut}^{\rm tag} < 0.1 \, \ebeam $ and  
 $ E_{\rm neut}^{\rm sig} < 0.1 \, \ebeam $,
 while the $(\mu,e)$, $(e,\mu)$ and
 $(\mu,\mu)$ modes require 
 $ E_{\rm neut}^{\rm tag} < 0.2 \, \ebeam $ and  
 $ E_{\rm neut}^{\rm sig} < 0.2 \, \ebeam $. 
The $(e,X)$, $(\mu,X)$ and $(X,X)$ modes all require that 
 $ E_{\rm neut}^{\rm tag} < 0.4 \, \ebeam $ and  
 $ E_{\rm neut}^{\rm sig} <  0.4 \, \ebeam $. In addition the $(e,X)$ mode
 requires fewer than 4 unassociated showers with energies above
 $0.4 \, \ebeam$.

For the  $(e,e)$ and $(\mu,\mu)$ modes, 
contamination from radiative dileptons  is reduced by requiring
 $P_{\mathrm{tag}}+P_{\mathrm{sig}} < 1.5 \,\ebeam $.
To reduce backgrounds from $e^+e^- \to {\it l}^+  {\it l}^- 
\gamma \gamma$ and 
$e^+e^- \to e^+e^- {\it l}^+ {\it l}^-$ in the $(e,e)$, $(\mu,\mu)$, 
$(X,X)$ categories, the minimum polar angle
 of any unseen particles, deduced from energy-momentum conservation,
 is required to point into the barrel region, where 
 calorimetry cuts will ensure rejection.

\begin{figure}
\includegraphics*[width=3.2in]{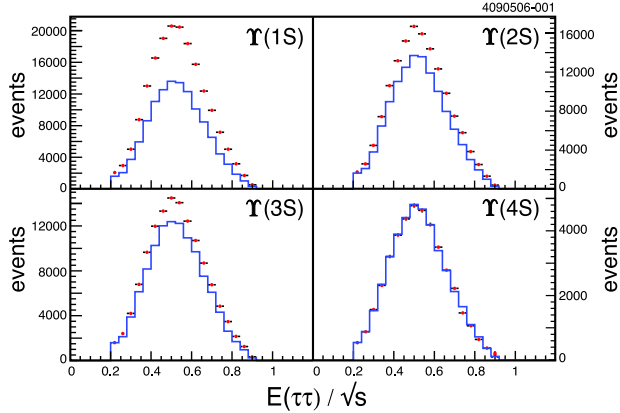}
\caption{ Total energy distributions, scaled to center of mass energy, 
$E_{\tau\tau}/\roots$
for the $\tau^+\tau^-$ final states
 at the $\Upsilon(nS), n=1,2,3,4$ on (points) and scaled off (histograms) resonance. 
The excess of on-resonance relative to scaled off-resonance data 
 is attributed to $\Upsilon(nS), n=1,2,3$ decays.
 The  agreement
 at the $\Upsilon(4S)$ tests the validity of the subtraction technique.
}
\label{fig:onoff}
\end{figure}

Figure~\ref{fig:onoff} shows the superimposed on-resonance  and scaled 
off-resonance total energy distributions
 for the $\tau^+\tau^-$ sample for all resonances. The scale factor is 
${\cal S} =  ( {\cal L}_{\mathrm{On}}/{\cal L}_{\mathrm{Off}})
            ( s_{\mathrm{Off}}/ s_{\mathrm{On}}) \delta_{\mathrm{interf}}$, 
where
 ${\cal L}$ and $s$ are the data luminosity and squared center of mass 
 collision energies on and off the resonances, and $\delta_{\mathrm{interf}}$
 is an interference correction.
  The luminosity  is derived from the  process 
$e^+e^-\to\gamma\gamma$ \cite{lumi}, which does not suffer backgrounds 
from direct $\Upsilon$ decays. The interference  correction 
$\delta_{\mathrm{interf}}$ accounts for the small interference between
the process $e^+e^- \to {\it l l}$ and $e^+e^- \to \Upsilon \to {\it l l}$ 
and is estimated~\cite{idanko} to be 0.984 (0.961, 0.982) at the 
$\Upsilon(1S)$ $(2S,\, 3S)$ and negligible for the $\Upsilon(4S)$. The 
 interference largely cancels in the ratios considered in this Letter.
 The agreement of the distributions for the $\Upsilon(4S)$, which  also 
 extends to individual sub-samples, validates the subtraction technique, and 
 highlights the absence of any process  whose cross-section does not 
 vary as $1/s$. This agreement also indicates that no Monte Carlo simulation
 is needed to subtract continuum production from $\Upsilon(nS)$ data.

The ratio of measured relative lepton pair 
 production cross-sections at the off-resonance points, $\myratiooff = \sigma_{\tau\tau}/\sigma_{\mu\mu} $,
 with respect to that theoretically expected,
 $\myratiooffexp $,
  is 
 $\myratiooff/\myratiooffexp = 0.96\pm 0.03$ ($0.97 \pm 0.03 $, 
 $0.97 \pm 0.03 $, $1.00 \pm 0.03 $) below the $\Upsilon(1S)$ 
 ($2S$, $3S$, $4S$) for the sum of all
 $\tau$ decay mode pairs, with statistical and systematic uncertainties 
 added in quadrature.
 The expectation
 $\myratiooffexp= 0.83 \pm 0.02 \,\mathrm{(syst)} $~\cite{MyNote}
 is found
 to be numerically independent of the particular resonance considered.
 The
 reconstruction efficiencies are derived 
from the FPair~\cite{FPair} and 
Koralb/Tauola~\cite{koralb,tauola,photos} Monte Carlo simulations. 
 Backgrounds were 
corrected by using $e^+e^- \to q \overline{q} (q=u,d,c,s)$  
simulations~\cite{QQ, Jetset,photos, Geant}. 
 The scatter in the central values of $\myratiooff/\myratiooffexp$ indicates
 that systematic uncertainties are small.

The reconstruction efficiency for observing $\Upsilon \to \mu^+\mu^-$ is derived 
from the CLEO  {\textsc GEANT}-based simulation~\cite{QQ,Jetset,photos,Geant},
 as shown in Table~\ref{tab:NumberOfEvents}.
This efficiency is found to be constant
across the resonances.  
  The reconstruction efficiency for observing $\Upsilon \to \tau^+\tau^-$ is 
derived using the Koralb/Tauola event generator integrated 
into the detector simulation. Although this generator
  models  the process $e^+e^- \to \gamma^* \to \tau^+\tau^-$, the 
quantum numbers of the $\Upsilon$ and $\gamma$ are the same so
it can be used as long as initial state radiation (ISR) effects are not 
included. This efficiency is found to be consistent across all
 resonances within any given $\tau^+\tau^-$ decay channel.

  Results of the subtraction are summarized in Table~\ref{tab:NumberOfEvents},
 showing 
the first observation of $\Upsilon(3S)  \to \tau^+\tau^-$.

  \begin{table}[htb]
  \vspace{0.4cm}
 \begin{center}
 \begin{tabular}{ c  c  c  c }
 \hline 
 	&
 $\Upsilon(1S)$ \,\,&
 $\Upsilon(2S)$ \,\,&
 $\Upsilon(3S)$ \\ \hline
 \rule[-0.5mm]{0mm}{4mm}
$ \tilde{N} ( \mu^+\mu^-)$ ($10^3$)&
 $345 \pm 7$ \,\,&
 $121\pm 7$ \,\,&
 $82 \pm 7$ \\ 
 $ \epsilon(\mu^+\mu^-)$ ($\%$) &
  $65.4 \pm 1.2 $ \,\,&
  $65.0 \pm 1.1 $ \,\,&
  $65.1 \pm 1.2 $ \\ 
  $N(\Upsilon \to \mu^+\mu^-) $ ($10^3$) &
   $527 \pm 15 $ \,\,& 
   $185 \pm 11 $ \,\,&
    $ 126 \pm 11 $ \\ \hline 
 \rule[-0.5mm]{0mm}{4mm}
 $ \tilde{N} (\tau^+\tau^-) $ ($10^3$)&
 $60.1 \pm 1.5 $ \,\,&
 $21.8\pm 1.5 $ \,\,&
 $14.8 \pm 1.5 $ \\ 
 $ \epsilon(\tau^+\tau^-)$ ($\%$) &
  $11.2 \pm 0.1  $ \,\,&
  $11.3 \pm 0.1 $ \,\,&
  $11.1 \pm 0.1 $ \\ 
  $N(\Upsilon \to \tau^+\tau^-) $ ($10^3$) &
   $537 \pm 14 $ \,\,& 
   $193 \pm 12 $ \,\,&
    $ 132 \pm 13 $ \\ \hline 
\end{tabular}
\caption{ Events yields for $\Upsilon \to \mu^+\mu^-$ (top)
 and  $\Upsilon \to \tau^+\tau^-$ (bottom). 
 Shown are the number of events  ($\tilde{N}_{\it l l}$) after subtraction of backgrounds estimated
  from scaled off-resonance data and $\Upsilon$ feed-through estimated from the
   Monte Carlo simulation,
 the signal efficiency ($\epsilon({\it l l})$), and 
 the total efficiency corrected number of signal events $N(\Upsilon \to {\it ll})=\tilde{N}_{\it l l}/\epsilon({\it l l} )$.
 The $\tau^+\tau^-$ events are summed over all $\tau$ decay modes.
 Uncertainties include data and Monte Carlo statistical 
uncertainties, uncertainties on backgrounds, and detector modeling (included only
 for $\epsilon(\mu^+\mu^-)$ to avoid double counting in the final ratio).
\label{tab:NumberOfEvents} }
\end{center}
\end{table}

 Backgrounds resulting from  
 cascade decays within the $b\overline{b}$ system to ${\it l l}$ are
  estimated using the Monte Carlo simulation, 
  with branching fractions scaled to the values measured in this study.
 Cascade backgrounds with
 non-$\mu^+\mu^-$ and non-$\tau^+\tau^-$ final states are estimated directly from 
 the  Monte Carlo simulation.

\begin{figure}
\includegraphics*[width=3.2in]{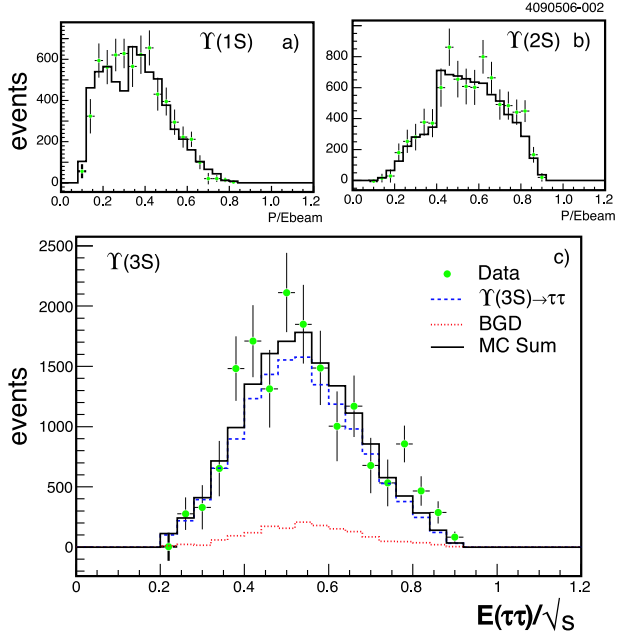}
\caption{ 
 Distributions for the $\tau^+\tau^-$ final states
 at the $\Upsilon(nS), n=1,2,3$ after subtraction of ${\cal S}$-scaled off-resonance
 data. The solid line shows the expected total signal and background
 distributions, assuming lepton universality. 
a)  $P_{\mathrm{sig}}/\ebeam$
 in $\Upsilon(1S)$ decays for the sum of $\tau$ decay 
 modes including exactly
 two identified leptons, 
b)   $P_{\mathrm{tag}}/\ebeam$
 in $\Upsilon(2S)$ decays for 
 $\tau$ modes including exactly one identified lepton,
c)  $E_{\tau\tau}/\roots$  for $\Upsilon(3S)$ for
 the sum of all $\tau$ modes, where signal and
  total background distributions are explicitly displayed.
 Uncertainties are statistical. The steps observed in a) and b) 
are due to momentum criteria in muon identification. 
}
\label{fig:signal}
\end{figure}

Figure~\ref{fig:signal} displays the off-resonance subtracted data, 
superimposed on Monte Carlo expectations.
The distributions shown are $P_{\mathrm{tag}}$, $P_{\mathrm{sig}}$ and $E_{\tau\tau}$ for 
 a sampling of $\tau$ decay modes. In all cases the Monte Carlo expectation
is consistent with the
 data assuming lepton universality and branching fractions from
 ~\cite{idanko}. The agreement across the various
 kinematic quantities indicates that backgrounds are well controlled.

\begin{figure}
\includegraphics*[width=3.0in]{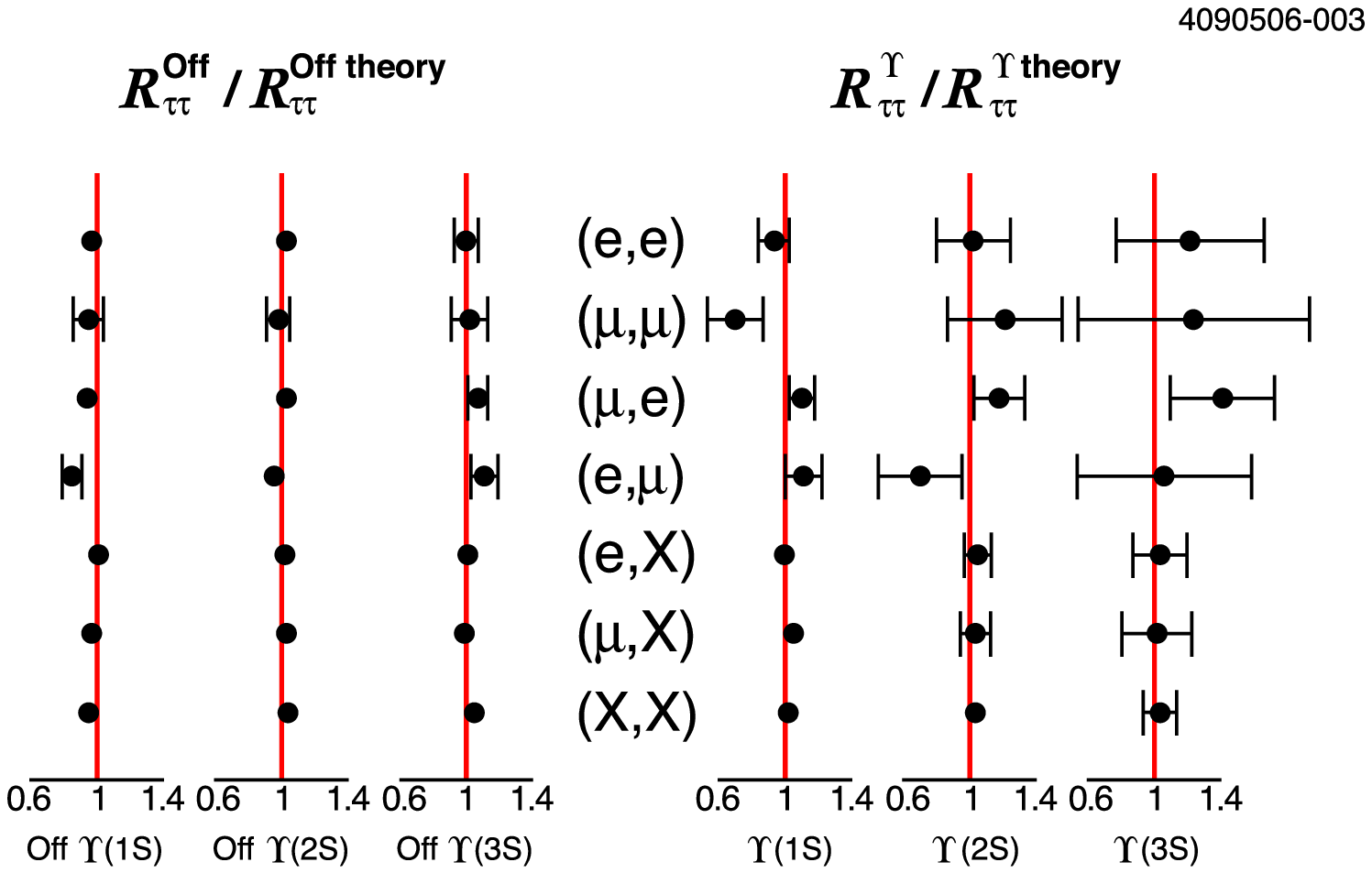}
\caption{ Breakdown by mode of off- and on-resonance data 
 at the different resonances.
 On the left, the ratio of the production cross section for 
 $e^+e^- \to {\it l}{\it l} ({\it l} = \tau,\mu)$,
 relative to its expectation,
 is plotted for data taken below the $\Upsilon$ for 
 each $\tau $ decay mode pair. On the right, the ratio of branching
 fractions for the process $\Upsilon(nS)\to {\it l}{\it l } ({\it l} = \tau,\mu, {\rm n=1,2,3} )$ relative to the expectation $\myratioexp = 1$
 is displayed.
The lines represent the Standard Model expectation. Errors
 shown are statistical.
}
\label{fig:breakdown}
\end{figure}

Figure~\ref{fig:breakdown} shows the agreement across all $\tau^+\tau^-$ 
 sub-samples of the ratio of off-resonance cross sections for $\tau^+\tau^-$ and
 $\mu^+\mu^-$ production, relative to expectation, as well as the ratio of 
branching fractions for each
 of these decay modes at the different $\Upsilon$ resonances, relative to the expectation $\myratioexp = 1$.
 The agreement across $\tau^+\tau^-$ sub-samples both on and off the resonances is
 again an indication that backgrounds are small and well estimated.

The ratio of branching fractions and final branching fractions are listed in 
 Table~\ref{tab:FinalResults}. These results show that lepton universality
 is respected  in $\Upsilon$ decay 
 within the $\approx 10\%$ measurement uncertainties.

  \begin{table}[htb]
  \vspace{0.4cm}
 \begin{center}
 \begin{tabular}{ c  c   c }
 \hline
 \rule[-0.5mm]{0mm}{4mm}
      \,\,\,\,   &
 $\myratio$ \,\,\,\, &
 $B(\Upsilon \to \tau^+ \tau^-  )$ ($\%$) 	\\  \hline
 $\Upsilon(1S)$ \,\,\,\, &
 $1.02 \pm 0.02 \pm 0.05$ \,\,\,\, &
 $2.54 \pm 0.04 \pm 0.12 $ \\
 $\Upsilon(2S)$ \,\,\,\,  &
  $1.04 \pm 0.04 \pm 0.05$ \,\,\,\, &
 $2.11 \pm 0.07 \pm 0.13 $ \\
 $\Upsilon(3S)$ \,\,\,\, &
   $1.05 \pm 0.08 \pm 0.05$ \,\,\,\, & 
   $2.52 \pm 0.19 \pm 0.15 $  \\ \hline
\end{tabular}
\caption{ 
 Final results on the ratio of branching fractions to $\tau^+\tau^-$ and
 $\mu^+\mu^-$ final states, and the absolute branching fraction for
 $\Upsilon \to \tau^+\tau^-$. Included are both statistical and systematic
 uncertainties, as detailed in the text. Results from Ref~\cite{idanko} 
 are used in deriving the final absolute branching fractions.
\label{tab:FinalResults} }
\end{center}
\end{table}

 Systematic uncertainties, summarized in Table~\ref{tab:syst}, are estimated 
 for the ratio of branching fractions, 
and for the absolute branching fraction. The ratio 
is insensitive to some common systematic uncertainties. 
 
 \begin{table}[htb]
 \vspace{0.4cm}
 \begin{center}
 \begin{tabular}{ c  c }
 \hline
 Source        &
 $\sigma_{\mathrm{syst}} $ ($\%$) \\ \hline
 ${\cal S }$ &             $0.2$ / $0.4$ / $0.3$ \\ 
 Background &              $0.1$ / $2.4$ / $1.3$ \\ \hline 
 $\tau$, $\mu$ Selection & $2.9$ / $2.9$ /  $2.9$\\ 
 $\Upsilon \to \mu^+\mu^-$ Model & $2.0$ / $2.0$ / $2.0$ \\ 
 $\Upsilon \to \tau^+\tau^-$ Model & $2.0$ / $2.0$ / $2.0$ \\ 
 Detector Model & $1.7$ / $1.7$ / $1.7$\\ 
 MC Statistics & $1.9$ / $1.0$ / $1.0$ \\ \hline 
 \rule[-0.5mm]{0mm}{4mm}
 $\sigma(\myratio)/\myratio$ & $4.8$ / $4.4$ / $4.6$ \\ \hline 
  $\sigma(B_{\tau\tau})/B_{\tau\tau}$  & $4.0$ / $3.8$ / $3.9$ \\ \hline
\end{tabular}
\caption{ 
Summary of systematic uncertainties for the $\Upsilon(nS),\, n=1/2/3$.
 The entry $\sigma(\myratio)/\myratio$  is the relative uncertainty on 
$\myratio$, while  $\sigma(B_{\tau\tau})/B_{\tau\tau}$ 
 indicates  uncertainties
 specific to $\tau$ decay modes used in addition to
 those in \cite{idanko} to obtain $B(\Upsilon \to \tau^+\tau^-)$.
 The uncertainty on ${\cal S}$ and the background are included in the 
 statistical  uncertainty as they depend chiefly on measurements
 made in this Letter.
 \label{tab:syst} }
\end{center}
\end{table}

 Most systematic uncertainties due specifically to ${\it l}{\it l}$ selection
are 
 derived by a variation of the selection criteria
 over reasonable ranges in the $\Upsilon(1S)$ sample, which has
 the lowest energy released in its decay.
The most significant of these are due to 
momentum selection ($1.3\%$), calorimeter energy selection ($1.1\%$) and 
angular selection ($1.1\%$).
  The systematic uncertainty due to modeling of the trigger is estimated 
to be $1.6\%$.

Backgrounds are assumed to be from $\Upsilon$ decays,  
chiefly due to cascade decays to lower resonances, and are estimated
 to be $2.5\%$ ($15\%$, $11 \%$) of the $\tau^+\tau^-$ sample at the 
$\Upsilon(1S) ( 2S, 3S)$, with
 an estimated uncertainty contribution to  $\myratio$ of $0.1\%$ ($2.4\%$, $1.3\%$).
 
The uncertainty due to detector modeling
 in \cite{idanko} was estimated to be $1.7\%$: this 
value is used
   here conservatively for the systematic uncertainty on the ratio. 
  
  The modeling of the physics in $\Upsilon(1S) \to \mu^+\mu^-$,
 obtained by varying the decay model for
 $\Upsilon \to \mu^+\mu^-$  between the Monte Carlo simulation
  and Koralb with ISR simulation turned off,
contributes a $2\%$ uncertainty.
This 
is  consistent with the variation 
in the product $\epsilon(\mu^+\mu^-)\sigma_{ee \to \mu\mu}$
using the FPair, 
     Koralb, and Babayaga~\cite{Babayaga} Monte Carlo simulations, and is thus 
likely conservative, 
as direct $\mu^+\mu^-$ production from
the $\Upsilon$ at the peak involves much lower energy final state
 photons than off resonance production.
An uncorrelated uncertainty of 
$2\%$ for modeling of $\Upsilon \to \tau^+\tau^-$ is assumed,
consistent with previous analyses, and is again  conservative as 
     on-resonance production of $\tau^+\tau^-$ final states involves fewer 
photons than direct      continuum production. 
To test the sensitivity to ISR simulation, the reconstruction efficiency for events
with no ISR simulation is compared to that for events generated
with ISR simulation turned on
and re-weighted according to the relative value of the $\Upsilon$ line
 shape at the $\tau$ pair mass. 
 These efficiencies agree to within $0.8\%$.

 The existence of a CP-odd Higgs boson,  $A^0$, with a mass near the
 $\Upsilon(1S)$, could induce
 a value of $\myratio$ not equal to one through its participation 
  in the decay chain 
 $\Upsilon(1S) \to \eta_b \gamma, \eta_b \to A^0 \to \tau^+\tau^-$,
 as detailed in  \cite{miguel}.
 By assuming that the
 mass of the $\eta_b(1S)$ is $100 \,\mathrm{MeV/c}^2$ below the $\Upsilon(1S)$ 
mass, consistent with the largest value in \cite{Rosner},
 the value quoted for $\myratio (1S)$ can be translated into an upper limit
 on the combined branching fraction of $B(\Upsilon(1S) \to \eta_b \gamma)
 B(\eta_b \to A^0  \to \tau \tau) < 0.27\% $ at $95\%$ confidence level,
 including systematic uncertainties.
Since the transition photon is not explicitly reconstructed,
this limit is valid for all $\eta_b$ that approximately
 satisfy 
 $( M(\Upsilon(1S)) - M(\eta_b) + \Gamma(\eta_b)) < 
{\cal O } (100 \,\mathrm{MeV/}c^2)$.

 In summary, using the full sample of on-resonance $\Upsilon(nS)$, $n=1,2,3$
CLEOIII data, we have made the first observation of the
 decay $\Upsilon(3S) \to \tau^+\tau^-$ and precision measurement of its branching
 fraction. We have reported the ratio
 of branching fractions of $\Upsilon$ decays to $\tau^+\tau^-$ and $\mu^+\mu^-$
 final states, and find these to be consistent with expectations from the
 Standard Model. These ratios have been combined with results from 
 \cite{idanko} to provide absolute branching fractions for the process 
 $\Upsilon \to \tau^+\tau^-$,
 resulting in the most precise single measurement of 
 $B(\Upsilon(1S)\to\tau^+\tau^-)$~\cite{pdg}, and
 a much improved value of $B(\Upsilon(2S)\to\tau^+\tau^-)$.
  The ratio of branching fractions for $\tau^+\tau^-$ and $\mu^+\mu^-$ final states
 has also been used to set a limit on a possible Higgs mediated decay window.

We gratefully acknowledge the effort of the CESR staff in providing us with
 excellent luminosity and running conditions. 
We also thank M.A.~Sanchis-Lozano for many illuminating discussions.
This work was supported by 
the A.P.~Sloan Foundation,
the National Science Foundation,
the U.S. Department of Energy, and
the Natural Sciences and Engineering Research Council of Canada.




\begin{thebibliography}{99}
\bibitem{MyNote}
 This expectation is lower than 1 because of the larger phase space available for  initial state radiation production of $\mu^+\mu^-$ relative to $\tau^+\tau^-$ final states.
\end{thebibliography}
\end{document}